\documentclass[a4paper, pra, amsmath,
 superscriptaddress, preprintnumbers, twocolumn ]{revtex4}

\usepackage{times,mathptm}
\usepackage{graphicx, color}

\begin{document}

\title{Cruising through molecular bound state manifolds with radio frequency}

\author{F. Lang}
\affiliation{Institut f\"ur Experimentalphysik und Forschungszentrum
f\"ur Quantenphysik, Universit\"at
  Innsbruck, 6020 Innsbruck, Austria}

\author{P. v. d. Straten}
\affiliation{Debye Institute, Universiteit Utrecht,
  3508 TA Utrecht, Netherlands}

\author{B. Brandst\"atter}
\affiliation{Institut f\"ur Experimentalphysik und Forschungszentrum
f\"ur Quantenphysik, Universit\"at
  Innsbruck, 6020 Innsbruck, Austria}

\author{G. Thalhammer}
\affiliation{Institut f\"ur Experimentalphysik und Forschungszentrum
f\"ur Quantenphysik, Universit\"at
  Innsbruck, 6020 Innsbruck, Austria}

\author{K. Winkler}
\affiliation{Institut f\"ur Experimentalphysik und Forschungszentrum
f\"ur Quantenphysik, Universit\"at
  Innsbruck, 6020 Innsbruck, Austria}

\author{P. S. Julienne}
\affiliation{Atomic Physics Division and Joint Quantum Institute,
National Institute of Standards and Technology, Gaithersburg, MD
20899, USA}

\author{R. Grimm}
\affiliation{Institut f\"ur Experimentalphysik und Forschungszentrum
f\"ur Quantenphysik, Universit\"at
  Innsbruck, 6020 Innsbruck, Austria}
\affiliation{Institut f\"ur Quantenoptik und Quanteninformation,
\"Osterreichische Akademie der Wissenschaften, 6020 Innsbruck,
Austria}

\author{J. Hecker Denschlag$^*$}
\affiliation{Institut f\"ur Experimentalphysik und Forschungszentrum
f\"ur Quantenphysik, Universit\"at
  Innsbruck, 6020 Innsbruck, Austria}

\date{\today}
\maketitle

{ The emerging field of ultracold molecules with their rich internal
structure is currently attracting a lot of interest. Various methods
~\cite{Doy04,Jon06,Koh06} have been developed to produce ultracold
molecules in pre-set quantum states. For future experiments it will
be important to efficiently transfer these molecules from their
initial quantum state to other quantum states of interest. Optical
Raman schemes \cite{Win07,Ber98} are excellent tools for transfer,
but can be involved in terms of equipment, laser stabilization and
finding the right transitions. Here we demonstrate a very general
and simple way for transfer of molecules from one quantum state to a
neighboring quantum state with better than 99\% efficiency. The
scheme is based on Zeeman tuning the molecular state to avoided
level crossings where radio-frequency transitions can then be
carried out. By repeating this process at different crossings,
molecules can be successively transported through a large manifold
of quantum states. As an important spin-off of our experiments, we
demonstrate a high-precision spectroscopy method for investigating
level crossings. }

Using radio-frequency (rf) for molecular spectroscopy is an
established technology dating back as far as the early 20th century
\cite{Ram56,Dem88}. In ultracold molecular gases it was employed in
recent years to measure binding energies of weakly bound states
\cite{Reg03,Chi04,Bar05,Koh05,Chi05a,Sch07} or to produce ultracold
molecules by associating ultracold atoms near a Feshbach resonance
\cite{Tho05,Osp06,Ber06}. Transferring ground state molecules
between states of different vibrational quantum numbers using rf as
demonstrated in this letter is not obvious. The spatial
wavefunctions of different vibrational states are in general
orthogonal to each other, trivially leading to a vanishing
transition matrix element for magnetic dipole transitions. However,
even simple molecules like Rb$_2$ have a complex level structure,
\begin{figure}[h]
\includegraphics[width=\columnwidth]{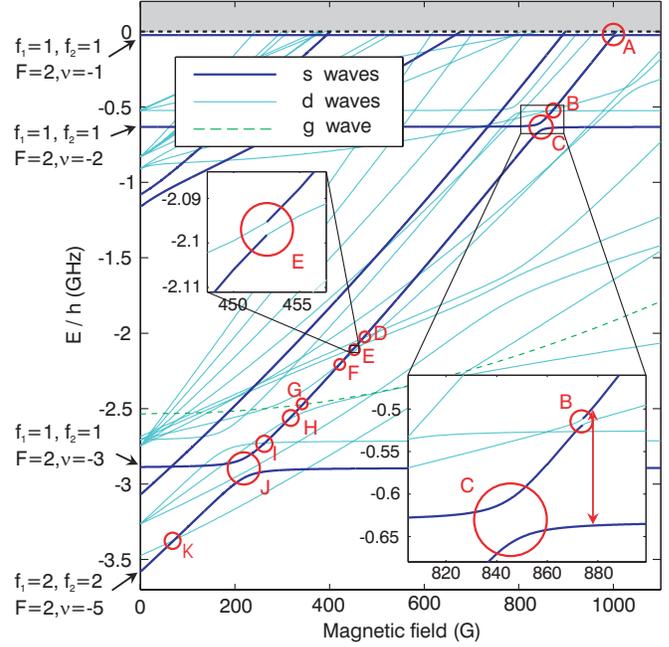}
\caption{ {\bf Path through manifold of molecular levels.} Energy
spectrum of relevant molecular levels of $^{87}$Rb$_2$  in the
electronic ground state with $m_{\text{Ftot}} = 2$. The zero of
energy is taken to be that of two separated atoms at each field
strength. Molecules are transported through the bound level manifold
by traversing avoided crossings, marked A - K. Out of the number of
existing $g$-wave levels, we only show the single relevant one.
(1\,G = 0.1\,mT).} \label{fig:panel1_1}
\end{figure}
especially when weakly bound, mainly due to exchange interaction,
hyperfine structure and the Zeeman effect. As a consequence, levels
with different vibrational quantum numbers can intersect and mix,
leading to allowed rf transitions between them.

We perform our experiments with an ensemble of $2 \times 10^4$
ultracold $^{87}$Rb$_2$ molecules which are held in
 the vibrational ground states of
 micro traps formed by a cubic 3D optical lattice with
lattice period $415\,$nm ~\cite{Tha06}. There is no more than a
single molecule per micro trap and the lattice potential is deep
enough ($\approx$ 10$\mu$K $\times k_B$ ) to effectively isolate the
molecules from each other, shielding them from detrimental
collisions. The molecular ensemble is initially produced from an
atomic $^{87}$Rb Bose-Einstein condensate (BEC) after adiabatically
loading it into the lattice. Subsequent ramping over a Feshbach
resonance at 1007.4\,G \cite{Vol03} (1\,G = 0.1\,mT) produces
Feshbach molecules and a final purification step removes all
chemically unbound atoms \cite{Tha06}.

Figure~\ref{fig:panel1_1} shows the relevant molecular level
spectrum for our experiment as calculated by a coupled channel model
\cite{Dul95,Tie98} based on adjusted ab initio Rb$_2$
Born-Oppenheimer potentials \cite{Kra90}. The spectrum is located
just below the dissociation threshold for $f = 1, m_f = 1$ ground
state atoms. It essentially consists of  straight lines of {\em s-}
and {\em d-}wave levels (corresponding to a rotational angular
momentum $l = 0, 2$, respectively)
\begin{figure}[h]
\includegraphics[width=\columnwidth]{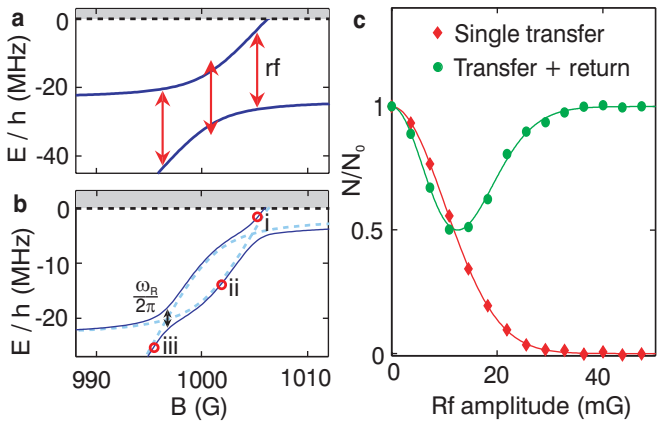}
\caption{ {\bf Adiabatic population transfer across avoided crossing
(ATAC) with radio frequency (rf)}. {\bf a}, Zoom into avoided
crossing A of Fig. 1. Transitions between the upper and lower branch
can be induced with rf. { \bf b}, Dressed state picture. Coupling
the lower and upper branch in { \bf a} with blue detuned rf induces
two avoided crossings with splitting $\omega_{R}$. Blue dashed
lines: case if $\omega_{R}$ = 0. For better visibility, the detuning
and $\omega_{R}$ are chosen larger than in the experiment. { \bf c},
Measured transfer efficiency for transfer from upper branch to lower
branch (red data) and back (green data). The solid lines are fits to
a Landau-Zener model. The rf amplitude is calibrated with a global
uncertainty of about 40\%. The ramp speed is 1.3\,G/ms. }
\label{fig:panel1_2}
\end{figure}
which intersect each other. These levels are further characterized
by their respective quantum numbers at zero magnetic field,
i.e.~global angular momentum $F_{\text{tot}}, m_{\text{Ftot}} = 2$,
the angular momenta $f_1, f_2$ of the atomic constituents, their
combined angular momentum $F$,  and the vibrational quantum number
$\nu$. In general, where two levels intersect, coupling between them
gives rise to an avoided crossing.

In the following we will use the level spectrum like a street map,
as the molecules move through the manifold of molecular bound states
by sweeping the magnetic field. When arriving at a level
intersection one can turn off or go straight, traversing the avoided
crossing. In principle, the avoided crossing can be jumped via a
fast magnetic field ramp \cite{Mar07a,Mar07b}. This, however, is
limited to very small splittings (typically $<$ 200 kHz $\times h$)
due to practical limitations of the controllable magnetic ramp
speed. This constraint can be easily overcome using an rf transition
as we demonstrate below.

As an example for cruising through molecular bound state levels we
choose the diagonal path in Fig. 1, as marked with the red circles A
to K, each indicating an avoided crossing. This converts our
Feshbach molecules with their weak binding energy of
24\,MHz$\times\,h$ to a deeper bound level at zero magnetic field,
3.6\,GHz$\times h$ below the $f=1, m_f = 1$ dissociation limit.

Fig.~\ref{fig:panel1_2}a shows an expanded view of the first avoided
crossing A. The upper branch is connected to the Feshbach resonance
at 1007.4\,G and is initially populated with Feshbach molecules. We
use adiabatic passage as a very efficient way for population
transfer to the lower branch.
\begin{figure}[h]
\includegraphics[width=\columnwidth]{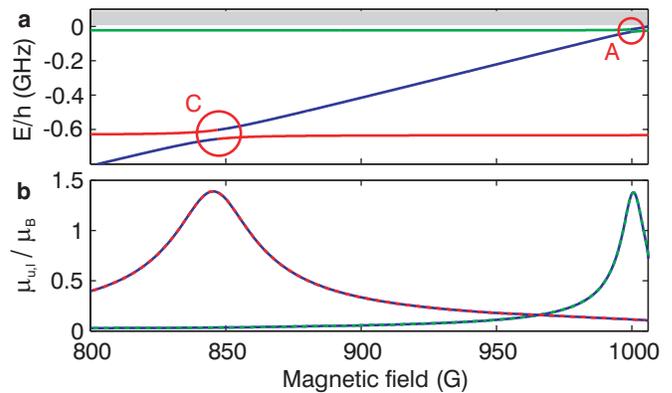}
\caption{ {\bf Calculated magnetic dipole matrix elements for the
avoided crossings A and C of Fig. 1}. Color coding identifies the
levels (see {\bf a}) with their respective transition (see {\bf b}).
} \label{fig:panel1_3}
\end{figure}
A longitudinal rf field (i.e. the rf magnetic field points in the
same direction as the magnetic bias field) couples upper and lower
branch via a magnetic dipole transition. We use a frequency of
13.6\,MHz, which is blue detuned with respect to the minimal
splitting of the avoided crossing. In the dressed state picture this
produces two rf induced avoided crossings, each with a splitting
corresponding to the Rabi frequency $\omega_R$, (see
Fig.~\ref{fig:panel1_2}b). At the beginning of the transfer the rf
field is off. The molecules which are initially located at the
position of the Feshbach resonance (point i in
Fig.~\ref{fig:panel1_2}b), are brought to point ii by lowering the
magnetic field to $B$= 1001 G. We then switch on the non-resonant rf
field and afterwards adiabatically lower the magnetic field $B$
until we reach point iii. Subsequent switching off the rf field
completes the transfer of the molecules to the lower branch.
Figure~\ref{fig:panel1_2}c shows the fraction $N/N_0$ of remaining
molecules in the upper branch after the transfer to the lower branch
(red data) as a function of the rf field amplitude $B_{\text{rf}}$.
For sufficiently high amplitude no more molecules are detected (see
Methods section for details on molecule detection). To verify that
molecules are not simply lost for high amplitude, we also carry out
a transfer back (iii $\rightarrow$ ii $\rightarrow$ i) to the upper
branch (green data). All molecules can be recouped again for strong
enough rf fields. From similar measurements with repeated transfers
between upper and lower branch we infer single transfer efficiencies
of up to 99.5\%. Our experimental data are well-fitted with the well
known Landau-Zener model (solid lines) \cite{Vit96} where the
transfer probability for a single transition is given by $ 1- \exp(-
\pi \omega_R^2 \hbar /2 | \dot{B}| \ |\mu_2- \mu_1| )$. Here
$|\dot{B}|$ is the ramp speed and $ \mu_1, \mu_2$ are the magnetic
moments of the two states.

After this successful demonstration at crossing A, we will use {\em
adiabatic transfer across an avoided crossing} (ATAC) repeatedly for
traversing the remaining crossings on our path. For these ATAC
transfers we typically apply radio-frequency with field amplitudes
$B_{\text{rf}}\sim50$\,mG, corresponding to Rabi frequencies
$\omega_{\text{rf}}\sim 2 \pi \times 70$\,kHz
 and ramp the magnetic bias field over about 1\,G in 1\,ms. We typically find
avoided crossings to lie within a few Gauss of their predicted
magnetic field position based on the coupled channel calculation.
This identification helps us also to verify that the molecules are
in the right quantum level during transport (see Methods section).

After traversing A, the next wide {\em s}-wave crossing is C. Before
we get to C, however, we hit the  avoided crossing B at 874\,G with
a $\sim$7\,MHz$\times\,h$ splitting, based on an intersecting {\em
d}-wave level  (see inset Fig.~\ref{fig:panel1_1}). To circumvent
crossing B we carry out the ATAC transfer between the {\em s}-waves
levels already at 876\,G, far from the {\em s}-wave crossing.

This raises the question, how far from an avoided crossing the rf
transitions can still be driven. Figure~\ref{fig:panel1_3}a is a
zoom into the energy spectrum showing the avoided crossings A and C.
Figure~\ref{fig:panel1_3}b plots the corresponding calculated
magnetic dipole matrix elements $\mu_{u,l}$ between the
corresponding upper and lower level branches using the coupled
channel model (see also Methods section).  The matrix elements are
clearly peaked at their respective crossing, reaching values of more
than a Bohr magneton. Such high coupling strengths are in agreement
with our measurements in Figure~\ref{fig:panel1_2}c where
$\mu_{u,l}$ can be extracted from the fits by using $\omega_R =
B_{\text{rf}} \mu_{u,l}$ and measuring $B_{\text{rf}}$. The width of
the peaks scales with the energy splitting of their avoided
crossing. When moving away from the crossing at $B_0$ the matrix
elements vanish inversely proportional to $ |B-B_0|$. A more
detailed discussion based on a simple model is presented in the
Methods section.

Continuing our path down by lowering the magnetic field we hit
consecutively five avoided crossings (D,E,F,H,I) with {\em d}-wave
states. The corresponding energy splittings are on the order of
1\,MHz$\times\,h$ and are each crossed by the ATAC method, which
demonstrates its universal character. In general, however, ATAC
transfers at narrow avoided crossings are technically more
challenging due to a small magnetic field range of strong coupling
and thus a greater susceptibility to magnetic field noise which can
lead to unwanted non-adiabatic transitions. Coupling to a {\em
g}-wave state is observed as well (crossing G), but it is weak
enough to be overcome by diabatic ramping of the magnetic field.
Finally, after crossing J and K we reach zero magnetic field, with
the molecules in state $|l = 0, F_{\text{tot}} = F= f_1 = f_2=
m_{\text{Ftot}} =2,\nu=-5\rangle$, 3.6 \,GHz below the $f_1 = f_2
=1, m_{f1} = m_{f2} = 1$ threshold.  We have also produced {\em
d}-wave molecules at zero magnetic field ($|l = 2,F_{\text{tot}} =2,
F=0, f_1 = f_2= m_{\text{Ftot}} = 2,\nu=-5 \rangle$) by
adiabatically following the upper branch in crossing K, i.e. taking
a right turn. The complete transfer down across all 10 avoided
crossings takes about 90\,ms with a global transfer efficiency of
about $~50\%$ (for molecule detection see Methods section).  The
losses during transfer can be explained mainly by the limited
molecular lifetime of 280\,ms in the lattice, due to inelastic
scattering of lattice photons \cite{Tha06}, and by not fully
optimized transfers at several crossings.

Besides the ATAC transfer of molecules between quantum levels, we
\begin{figure}[h]
\includegraphics[width=\columnwidth]{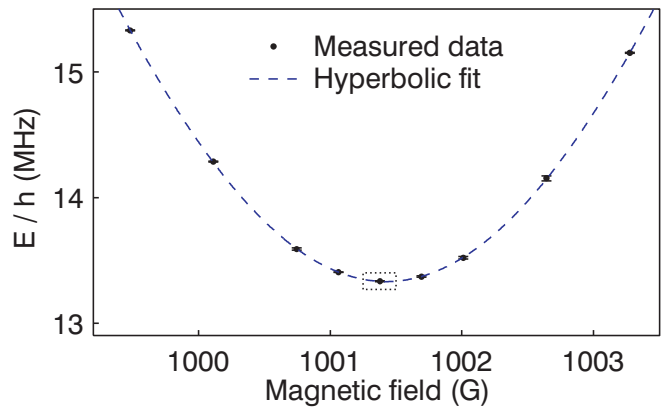}
\caption{ {\bf Spectroscopy of an avoided crossing}. For the avoided
crossing A the splitting is measured for various magnetic fields by
determining the resonant transition frequency. The error bars are
typically smaller than the size of the plot symbol. The dashed line
is a hyperbolic fit yielding a minimum frequency of ($13.331\pm0.005
$)\,MHz.} \label{fig:panel2_1}
\end{figure}
also developed a high precision spectroscopy method for measuring
the minimal energy splitting of an avoided crossing.  For a given
avoided crossing the energy splitting is measured for various
magnetic fields. We use two methods. Method 1 determines the
resonance frequency  for transfer of molecules between the two
branches of the avoided crossing. Using a single rf pulse of a few
ms length we look for the frequency of maximal transfer. The
corresponding data for crossing A are shown in
Fig.~\ref{fig:panel2_1}a and are very well fitted with a hyperbolic
curve, yielding a splitting of $(13.331\pm0.005)$\,MHz$\times h$.

\begin{figure}[h]
\includegraphics[width=\columnwidth]{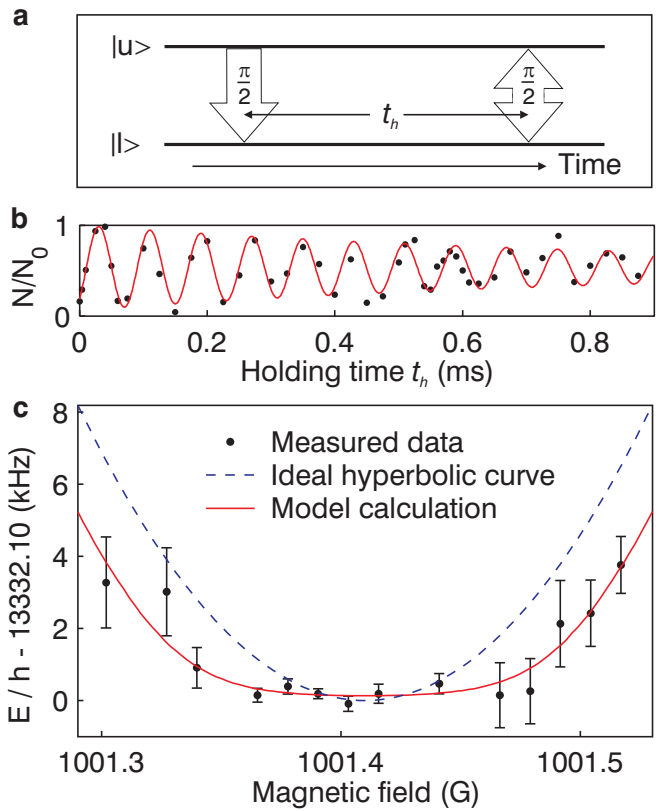}
\caption{ {\bf Spectroscopy with Ramsey interferometry}. {\bf a},
Ramsey scheme consisting of two $\frac{\pi}{2}$-pulses of rf and a
holding time $t_h$. {\bf b}, Fraction of remaining Feshbach
molecules after a Ramsey scan  as holding time is varied. $B$
=1001.39\,G. The oscillation frequency corresponds to the detuning
of the rf from the splitting. {\bf c}, Measured splittings, in a
narrow magnetic field region (also indicated by the small dotted box
in Fig. 4). The deviation from the hyperbolic curve (dashed blue
line) results from magnetic field fluctuations and inhomogeneities
and is reproduced by our model calculation (solid red line). }
\label{fig:panel2_2}
\end{figure}

In order to increase the precision we use Method 2, where we perform
a Ramsey-type interferometric measurement (see
Fig.~\ref{fig:panel2_2}a). A $\frac{\pi}{2}$-pulse of rf transfers
50\% of the Feshbach molecules to the lower branch, creating a 50/50
coherent superposition. After a hold time $t_h$ and a second
$\frac{\pi}{2}$-pulse the number of Feshbach molecules is detected.
We observe an oscillation of this population $N$ (see
Fig.~\ref{fig:panel2_2}b) which corresponds precisely to the
detuning of the rf field from resonance.
Coherence times reach 1ms. Besides its inherently high precision,
this method offers the advantage of being free from AC-Stark shifts
as the radio frequency is shut off during $t_h$. We verified this by
measuring $\omega_{res}$ for both positive and negative detuning,
giving the same results within the error margin.
Fig.~\ref{fig:panel2_2}c shows a set of data taken in the region
indicated by the dashed box in the center of
Fig.~\ref{fig:panel2_1}. The clear deviation from the hyperbolic
curve results from a $\sim$2 G/mm magnetic field gradient across the
molecular cloud ($\sim$20 $\mu$m diameter) in combination with
$\sim$20 mG fluctuations of the magnetic field during the time of a
scan. This behavior is reproduced by our model calculation (solid
red line) taking these experimental imperfections into account. From
the model we obtain a best estimate of the minimum splitting $
(13.33210\pm 0.00015)$\,MHz$\times h $ for the ideal hyperbolic
curve. The upshift of about 150 Hz of the minimum of the model curve
with respect to the hyperbolic curve is due to averaging over the
magnetic field inhomogeneities.

We have also performed detailed measurements of the energy splitting
at the avoided crossings marked C, E, and J in
Fig.~\ref{fig:panel1_1}. These data are shown in Table 1.
\begin{table}[h]
\label{tab:crossings}
 \begin{tabular}{|c|c|c|c|}
  \hline
  & partial waves & splitting/$h$ (MHz) & $B_{0}$ (G) \\
  \hline
  A & s-s & $13.33210\pm0.00015$  & $1001.4\pm0.2$ \\
  C & s-s  & $44.756\pm0.006$      & $845.8\pm0.2$ \\
  E & s-d  & $2.36\pm0.01$         & $466.1\pm0.2$ \\
  J & s-s  & $110.48\pm0.01$       & $218.8\pm0.2$ \\
  \hline
 \end{tabular} \medskip
\caption{
 Precision data for the minimal energy splitting of the
 avoided crossings A, C, E, and J
 of Fig. 1a. The second column indicates the partial waves
 involved in the crossing.
 The third column gives the minimal splitting which is located at
 magnetic field $B_0$. The splitting for A is measured with the Ramsey method, whereas
 the splittings for C, E and J were determined via Method 1 (resonant
 transfer).}
\end{table}
The measurements are complementary to conventional bound state
spectroscopy because instead of measuring the plain energy spectrum
of the bound states, our method determines the strength of the
coupling between levels. The precision of our data is several orders
of magnitude better than the accuracy of our current coupled channel
model, as well as the model of the Eindhoven group \cite{Kok07}.
Thus the data can serve to improve and test the theoretical models
used to calculate molecular energy levels.

To conclude, we have demonstrated an efficient and universal method
(ATAC) to transfer molecules between quantum states. This method is
based on a combination of rf pulses and magnetic field ramping and
can be applied to any molecule, since the only requirement is the
existence of magnetically tunable avoided crossings. This opens
interesting perspectives for experiments in cold
collisions~\cite{Chi05,Sta06,Zah06}, chemistry in the ultracold
regime, high resolution spectroscopy, and molecular BEC. In
particular, we plan to use the rf transfer method to prepare
Feshbach molecules in a convenient start position for an optical
Raman transition  to a deeply bound molecular state or even the
vibrational ground state \cite{Win07,Jak02}. In principle, the
triplet vibrational ground state for Rb$_2$ can also be reached
directly via the ATAC transfer scheme. By cruising back and forth
within a 5 Tesla range between consecutive avoided crossings one can
step down the vibrational ladder. Further, the ATAC scheme can be
extended in a straight forward manner to avoided crossings which are
tuned by electrical fields.

\section*{Methods}
\label{sec:methods}

{\bf Detection of molecules and their quantum state}

In order to detect the molecules at any stage during their transport
through the manifold of molecular levels, we trace back exactly the
path we have come before, adiabatically traversing all avoided
crossings in the opposite direction. We end up with Feshbach
molecules which are dissociated into unbound atoms by sweeping over
the Feshbach resonance at 1007.4~G. These atoms are then counted via
standard absorption imaging, after switching off the optical lattice
and the bias magnetic field.

We use two methods to verify that molecules are in the right quantum
level during transport. 1) Checking for consistency between
predicted and experimentally found avoided crossings, in  terms of
magnetic field location and energy splitting.  2) Optical
spectroscopy to measure the binding energy of molecules. By
irradiating the molecules with resonant laser light we transfer the
molecules to an electronically excited molecular level, $ |0_g^-,
\nu = 31, J = 0\rangle$ \cite{Win07}, leading to losses. The shift
of this laser frequency compared to the frequency of the
photoassociation transition to the same excited molecular level
corresponds to the binding energy of the molecules.

{\bf  Simple model for rf transitions at an avoided crossing}

A simple two level model gives insight into the mechanism for the rf
transitions at the avoided crossing. Two molecular bare levels
$|b1\rangle, |b2\rangle$ with magnetic moments $\mu_1$ and $\mu_2$
cross at a magnetic field $B = B_0$. The Hamiltonian for these
levels reads
\begin{eqnarray}
  \nonumber
  \hat{H}=
\left( B-B_0 + B_{\text{rf}} \cos(\omega_{\text{rf}}t) \right)
  \begin{pmatrix}
     \mu_1  & 0 \\
    0             & \mu_2 \\
  \end{pmatrix}
  +
{\hbar \over 2}
  \begin{pmatrix}
    0               & \Omega \\
    \Omega & 0 \\
  \end{pmatrix}.
 \label{equ:Htot}
\end{eqnarray}
  A coupling $\Omega$
between the two levels, e.g. due to exchange interaction or
dipole-dipole interaction, leads to mixing and the new eigenstates
$|u\rangle$ and $|l\rangle$. These states form the upper and lower
branch of an avoided crossing, similar to Fig. 2a. A longitudinal
magnetic rf field with amplitude $B_{\text{rf}}$ and frequency
$\omega_{\text{rf}}$ can drive transitions between levels
$|u\rangle$ and $|l\rangle$ which read
\begin{eqnarray}
 \label{equ:eigvec}
 \nonumber
 |u\rangle&=&\cos(\theta)|b1\rangle + \sin(\theta)|b2\rangle \\
 |l\rangle&=&-\sin(\theta)|b1\rangle + \cos(\theta)|b2\rangle,
\end{eqnarray}
with mixing angle $\theta = \arctan(\frac{\delta+\sqrt{\delta^2 +
\Omega^2}}{\Omega})$, where $\delta = (\mu_2 - \mu1)(B-B_0)$. The
matrix element for the rf transition is then
\begin{eqnarray}
 \mu_{u,l} &\equiv& \langle u |\begin{pmatrix}
     \mu_1  & 0 \\
    0             & \mu_2 \\
  \end{pmatrix}| l \rangle =
 (\mu_2 - \mu_1) \sin(2\theta) \\
 &=& 2 (\mu_2 - \mu_1) \frac{\Omega(\delta+\sqrt{\delta^2 +
\Omega^2})}{\Omega^2+(\delta+\sqrt{\delta^2 + \Omega^2})^2} .
\label{equ:matel}
\end{eqnarray}
Thus $\mu_{u,l}$ is resonantly peaked at the avoided crossing with a
width (FWHM) of $2 \sqrt{3} \Omega$ and vanishes as $1/(B-B_0)$ far
away from the crossing.

We find good agreement when comparing the matrix elements of our
simple model with the ones of the coupled channel model, given by
$\mu_{u,l} =  \ \langle u | \mu_B g_s S_z + \mu_N g_I I_z | l
\rangle$. Here $| u \rangle, | l \rangle$ are the wavefunctions as
calculated with the coupled channel model. $\mu_B, \mu_N$ are the
Bohr magneton and nuclear magneton, $g_s$ and $g_I$ are the
g-factors of the electrons and nuclei, respectively, and $S_z, I_z$
are the corresponding spin operator components in the direction of
the magnetic field.

\section*{Acknowledgements}

 We thank Wolfgang Ketterle, Cheng Chin, and Servaas
Kokkelmans for valuable discussions. This work was supported by the
Austrian Science Fund (FWF) within SFB 15 (project part 17). PvdS
acknowledges support within the ESF-program QUDEDIS during his stay
in Innsbruck. PSJ was partially supported by the US Office of Naval
Research.


\end{document}